\pdfoutput=1
\documentclass{article}
\usepackage[utf8]{inputenc}

\newcommand{\sltr}{\mathrm{SL}(2,\mathbb{R})}

\newcommand{\kbad}{\kappa_\text{sing}}

\newenvironment{eqaed}
    {
    \begin{equation}
        \begin{aligned}
    }
    {
        \end{aligned}
    \end{equation}\ignorespacesafterend
    }

\newcommand{\tablespacing}{{\vskip 1em}}

\usepackage{amsmath,amsthm,amssymb}
\usepackage[dvipsnames]{xcolor}
\usepackage{physics}
\usepackage{tensor}
\usepackage{graphicx}
\usepackage{float}
\usepackage{xcolor, colortbl}
\usepackage{hyperref}
\usepackage{authblk}
\usepackage{mathtools}
\usepackage{geometry}
\definecolor{light-gray}{gray}{0.9} 

\usepackage{setspace}
\usepackage{xparse}
\usepackage{esvect} 
\usepackage{slashed} 
\usepackage{dsfont} 
\usepackage[dvipsnames]{xcolor} 
\usepackage{empheq}
\usepackage{makecell}
\usepackage{subcaption}
\usepackage{float}
\usepackage{hyperref}
\hypersetup{
    colorlinks=true,
    linkcolor=blue,
    filecolor=magenta,      
    urlcolor=cyan,
}
\usepackage{import}
\usepackage{makeidx}
\title{ On  Perturbation Theory and Critical Exponents for Self-Similar Systems
}\author{Ehsan Hatefi\footnote{ehsan.hatefi@sns.it}} 
\author{Adrien Kuntz\footnote{ehsan.hatefi@sns.it}}
\affil{Scuola Normale Superiore and I.N.F.N,\protect\\ 
Piazza dei Cavalieri 7, 56126, Pisa, Italy\vspace{.7em}}

\begin{document}

\maketitle

\vspace{-0.7cm}


\begin{abstract}


Gravitational critical collapse in the Einstein-axion-dilaton system is known to lead to continuous self-similar solutions characterized by the Choptuik critical exponent $\gamma$. We complete the existing literature on the subject by computing the linear perturbation equations in the case where the axion-dilaton system assumes a parabolic form. Next, we solve the perturbation equations in a newly discovered self-similar solution in the hyperbolic case, which allows us to extract the Choptuik exponent. Our main result is that this exponent depends not only on the dimensions of spacetime but also the particular ansatz and the critical solutions that one started with.
\end{abstract}

\newpage

\section{Introduction}

A well-known property of black holes as the end state of gravitational collapse is that they are completely defined by only three numbers : their mass, angular momentum, and charge. Choptuik revealed in~\cite{Chop} that there may be a fourth universal quantity characterizing the collapse itself. Following the study of Christodolou in~\cite{Christodolou} on the spherically symmetric collapse of scalar fields, Choptuik discovered a critical behavior illustrating some sort of discrete spacetime self-similarity. Expressing the amplitude of the scalar field fluctuation by the number $p$, he found that $p$ should exceed a critical value $p_\text{crit}$ in order to form a black hole. Furthermore, for values of $p$ above the threshold, the mass of the black hole $M_\text{bh}$ (equivalently, its Schwarzschild radius $r_S$) exhibits the scaling law
%
%
%
%
\begin{equation}
r_S(p) \propto M_\text{bh}(p) \propto (p-p_\text{crit})^\gamma\,,
\end{equation}
where  the Choptuik exponent was found to be $\gamma\simeq 0.37$~\cite{Chop,Hamade:1995ce,Gundlach:2002sx} in 4d for a single real scalar field. Note that in general dimensions ($d \geq 4$) the definitions get changed~\cite{KHA,AlvarezGaume:2006dw}
\begin{equation}
 r_S(p) \propto (p-p_\text{crit})^\gamma \,, \quad M_\text{bh}(p) \sim (p-p_\text{crit})^{(D-3)\gamma}  \,.
\end{equation}

Along the same lines, diverse numerical simulations with different matter fields have been carried out\cite{Birukou:2002kk,Husain:2002nk,Sorkin:2005vz,Bland:2005vu,HirschmannEardley,Rocha:2018lmv}. For example the critical collapse of a perfect fluid was performed in~\cite{AlvarezGaume:2008qs,evanscoleman,KHA,MA}.
In \cite{evanscoleman} the authors found $\gamma \simeq 0.36$ and hence it was conjectured in \cite{Strominger:1993tt} that $\gamma$  may be universal for any matter field that is coupled to four dimensional gravity. Later on, in~\cite{KHA,MA,Hirschmann:1995qx} it was discovered that the Choptuik exponent can be explored by dealing with perturbations of the self-similar solutions. In order to do so, one needs to perturb any field $h$ (be it the metric or the matter content) as follows

\begin{equation}
    h = h_0 + \varepsilon \, h_{-\kappa}
\end{equation}

where the perturbation $h_{-\kappa}$ has scaling $-\kappa \in \mathbb{C}$ which labels the different modes. Among the allowed values of $\kappa$, 
we define the most relevant mode $\kappa^*$ as the highest value of $\Re(\kappa)$\footnote{The minus sign indicates a growing mode near the black-hole formation time $t \rightarrow 0$.}. It was shown in~\cite{KHA,MA,Hirschmann:1995qx} that $\kappa^*$ is related to the Choptuik exponent by

\begin{equation}
    \gamma = \frac{1}{\Re \kappa^*}\,.
\end{equation}

 In~\cite{AE} the case of  axial symmetry had been studied and the critical collapse in the presence of shock waves was reviewed in~\cite{AlvarezGaume:2008fx}. The case of axion-dilaton critical collapse coupled to gravity in four dimension was first examined in~\cite{Hirschmann_1997} which found the value $\gamma \simeq 0.2641$, hence raising serious doubts concerning the universality of $\gamma$ in four dimensions.

One motivation to study critical collapse in the axion-dilaton system is the AdS/CFT correspondence \cite{Maldacena:1997re}, relating Choptuik exponent, the imaginary part of quasinormal modes, and the dual conformal field theory \cite{Birmingham:2001hc}. Other motivations include the holographic description of black hole formation \cite{AlvarezGaume:2006dw}
as well as the physics of black holes and its applications \cite{Hatefi:2012bp}. In type IIB string theory one is often interested in exploring the gravitational collapse on spaces that can asymptotically approach to $AdS_5 \times S^5$ where the matter content is described by the axion-dilaton system and the self-dual 5-form field. 

The entire families of distinguishable continuous self similar solutions of the Einstein-axion-dilaton system in four and five dimensions for all the three conjugacy classes of $\sltr$ were recently explored in~\cite{ours} that generalized the previous efforts done in~\cite{AlvarezGaume:2011rk, hatefialvarez1307}. Based on some robust analytic and numerical techniques in~\cite{Antonelli:2019dqv}, 
we did perturb critical solution of four-dimensional elliptic critical collapse and were able to recover the known value~\cite{Hirschmann_1997} of $\gamma \sim 0.2641$. Hence this provides strong confidence in our ability to obtain the other critical exponents in different dimensions as well as for different classes of solutions.

In this article, after a brief recap on self-similar solutions to the Einstein-axion-dilaton system, we set up a linear perturbation analysis  which will allow us to extract the Choptuik exponent in any dimensions. The new methodology that we employ is quite generic and could be applied to other matter contents as well. Using this framework, we derive the perturbation equations in all conjugacy classes of $\sltr$, and particularly in the parabolic case which was not studied before. We extract the Choptuik exponent in a new branch of the 4d hyperbolic class of solutions and find that its value is different from the other branches of solutions. Thus, our results cast doubts concerning the universality of the Choptuik exponent.
\section{Self-Similar Solutions to Einstein-axion-dilaton configuration}\label{sec:unperturbed}

The Einstein-axion-dilaton system that coupled to gravity in $d$  dimensions is defined by the following action 

\begin{equation}
S = \int d^d x \sqrt{-g} \left( R - \frac{1}{2} \frac{ \partial_a \tau
\partial^a \bar{\tau}}{(\Im\tau)^2} \right) \; .
\label{eaction}
\end{equation}

 that can be described by the effective action of type II string theory \cite{Sen:1994fa,Schwarz:1994xn} where the axion-dilaton is defined by $\tau \equiv a + i e^{-\phi}$. This action enjoys the $\sltr$ symmetry
 \begin{equation} \label{eq:sltr}
     \tau \rightarrow M \tau \equiv \frac{\alpha \tau + \beta}{\gamma \tau + \delta} \; ,
 \end{equation}
 where $\alpha$, $\beta$, $\gamma$, $\delta$ are real parameters satisfying $\alpha \delta - \beta \gamma = 1$.
 It was known that once quantum effects are taken into account the $\sltr$ symmetry does reduce to $\mathrm{SL}(2,\mathbb{Z})$ and this S-duality was also believed to be a 
non-perturbative symmetry of IIB string theory~\cite{gsw,JOE,Font:1990gx}. Now from the above action one can read off the equations of motion 
\begin{equation}
\label{eq:efes}
R_{ab} = \tilde{T}_{ab} \equiv \frac{1}{4 (\Im\tau)^2} ( \partial_a \tau \partial_b
\bar{\tau} + \partial_a \bar{\tau} \partial_b \tau)\,,
\end{equation}
\begin{equation}\label{eq:taueom}
\nabla^a \nabla_a \tau +\frac{ i \nabla^a \tau \nabla_a \tau }{
\Im\tau} = 0 \,.
\end{equation}
We assume spherical symmetry on both background and perturbations so that the general form of the metric in $d$ dimensions is
\begin{equation}
    ds^2 = (1+u(t,r)) (-b(t,r)^2 dt^2 + dr^2) + r^2 d\Omega_q^2 \,,
\end{equation}
\begin{equation}
    \tau = \tau(t,r)\, , \quad q \equiv d-2 \; ,
\end{equation}
where $d\Omega_q^2$ is the angular part of the metric in $d$ spacetime dimensions.
A scale invariant solution is found by requiring that under a spacetime dilation (or scale transformation),  $ (t,r)\rightarrow ( \Lambda t,\Lambda r)$, the line element gets changed as $ ds^2 \rightarrow \Lambda^2 ds^2$. Thus, the metric functions should be scale invariant, i.e. $u(t,r) = u(z)$, $b(t,r) = b(z)$, $z \equiv -r/t$. Since the action~\eqref{eaction} is invariant under the $\sltr$ transformation~\eqref{eq:sltr}, 
$\tau$ only needs to be invariant and up to an $\sltr$ transformation, 
\begin{equation}\label{eq:tauscaling}
    \tau(\Lambda t, \Lambda r) = M(\Lambda) \tau(t,r)\,.
\end{equation}
We call a system of $(g,\tau)$ respecting the above properties a continuous self-similar (CSS) solution. Note that different cases do relate to various classes of $\eval{\dv{M}{\Lambda}}_{\Lambda=1}$~\cite{ours}, so that $\tau$ can take three different forms,

\begin{equation}\label{eq:tauansatz}
    \tau(t,r) = \begin{dcases}
                i \frac{1-(-t)^{i\omega}f(z)}{1+(-t)^{i\omega} f(z)}\,, & \quad \text{elliptic}\\[5pt]
                    f(z) + \omega \log(-t)\,, & \quad \text{parabolic}\\[5pt]
                    \frac{1-(-t)^{\omega}f(z)}{1+(-t)^{\omega} f(z)}\,,& \quad \text{hyperbolic}
                \end{dcases}
\end{equation}

where  $\omega$ is an unknown real parameter and the function $f(z)$ must satisfy $\abs{f(z)} < 1$ for the elliptic case and $\Im f(z)>0$ for the other two cases. Note that one can show that the the ansatz $\tau(t,r)= (-t)^\omega f(z)$ also leads to the same equations of motion for hyperbolic case (it is simply a conformal transformation of $\tau$). If we replace the CSS ans\"atze in the equations of motion we then get a differential system of equations for $u(z)$, $b(z)$, $f(z)$. Due to spherical symmetry one can show that $u(z)$ can be expressed in terms of $b(z)$ and $f(z)$ so that eventually we are left out with some ordinary differential equations (ODEs) 
\begin{align}\label{eq:unperturbedbp}
    b'(z) & = B(b(z),f(z),f'(z))\,, \\
    f''(z) & = F(b(z),f(z),f'(z))\,. \label{eq:unperturbedfpp}
\end{align}
The above equations have five singularities~\cite{AlvarezGaume:2011rk} located at $z = \pm 0$, $z = \infty$ and $z = z_\pm$ where the last singularities are defined by $b(z_\pm) = \pm z_\pm$. The latter correspond to the homotetic horizon and it can be shown that $z=z_+$ is just a mere coordinate singularity~\cite{Hirschmann_1997,AlvarezGaume:2011rk},  hence $\tau$ is regular across it which translates back to the finiteness of $f''(z)$ as $z\rightarrow z_+$. Now one may observe that the vanishing of the divergent part of $f''(z)$ gives us a complex valued constraint at $z_+$ which we denote by  $G(b(z_+), f(z_+), f'(z_+)) = 0$ where the explicit form of $G$ was given in \cite{ours}.

Using regularity at $z=0$ and some residual symmetries one obtains the initial conditions $b(0) = 1, f'(0) =0$

\begin{equation}
        f(0) = \left\{\begin{array}{l l l}
        x_0 & \text{elliptic}       & (0<x_0<1) \\
        i x_0 & \text{parabolic} & (0<x_0)\\
        1+i x_0 & \text{hyperbolic} & (0<x_0)
    \end{array}\right.
\end{equation}

Here $x_0$ is a real parameter. Hence, we have two constraints (the vanishing of the real and imaginary parts of $G$) and two parameters $(\omega,x_0)$. The discrete solutions in four and five dimensions were found in \cite{Antonelli:2019dqv}. These solutions are constructed by integrating numerically the equations of motion.
For instance, for the four dimensional elliptic case just one solution is determined \cite{Eardley:1995ns,AlvarezGaume:2011rk} as\begin{equation}
    \omega=1.176,\quad \abs{f(0)}=0.892,\quad z_+=2.605 \label{esi}
\end{equation}
To deal with self-similar solutions for parabolic cases, the following remarks are in order.
First, we have an additional  symmetry as follows

\begin{equation}\label{parabolic_rescaling}
    \omega \rightarrow K \omega\,,\quad f(z) \rightarrow K f(z)\,,\quad K \in \mathbb{R}_+
\end{equation}
then $\tau$ also transforms as
$\tau\rightarrow K \tau$, which means that if  $(\omega,\Im f(0))$ is a solution, so is $(K \omega,K\Im f(0))$. The reason behind it is that all equations of motion 
and the constraint $G(\omega,\Im f(0))$ are invariant under this new scaling. Therefore, the only real unknown parameter for parabolic class is the ratio $\omega/\Im f(0)$. We then need to look for
the zeroes of $G(\omega,\Im f(0))$ for only a real parameter $\omega/\Im f(0)$. Hence we just set $\Im f(0)=1$. For the five dimensional parabolic case, we draw below the plot of the zeroes of the real and imaginary parts of $G(\omega,1)$.
\begin{figure}[H]
    \centering
    \includegraphics[width=3.5in]{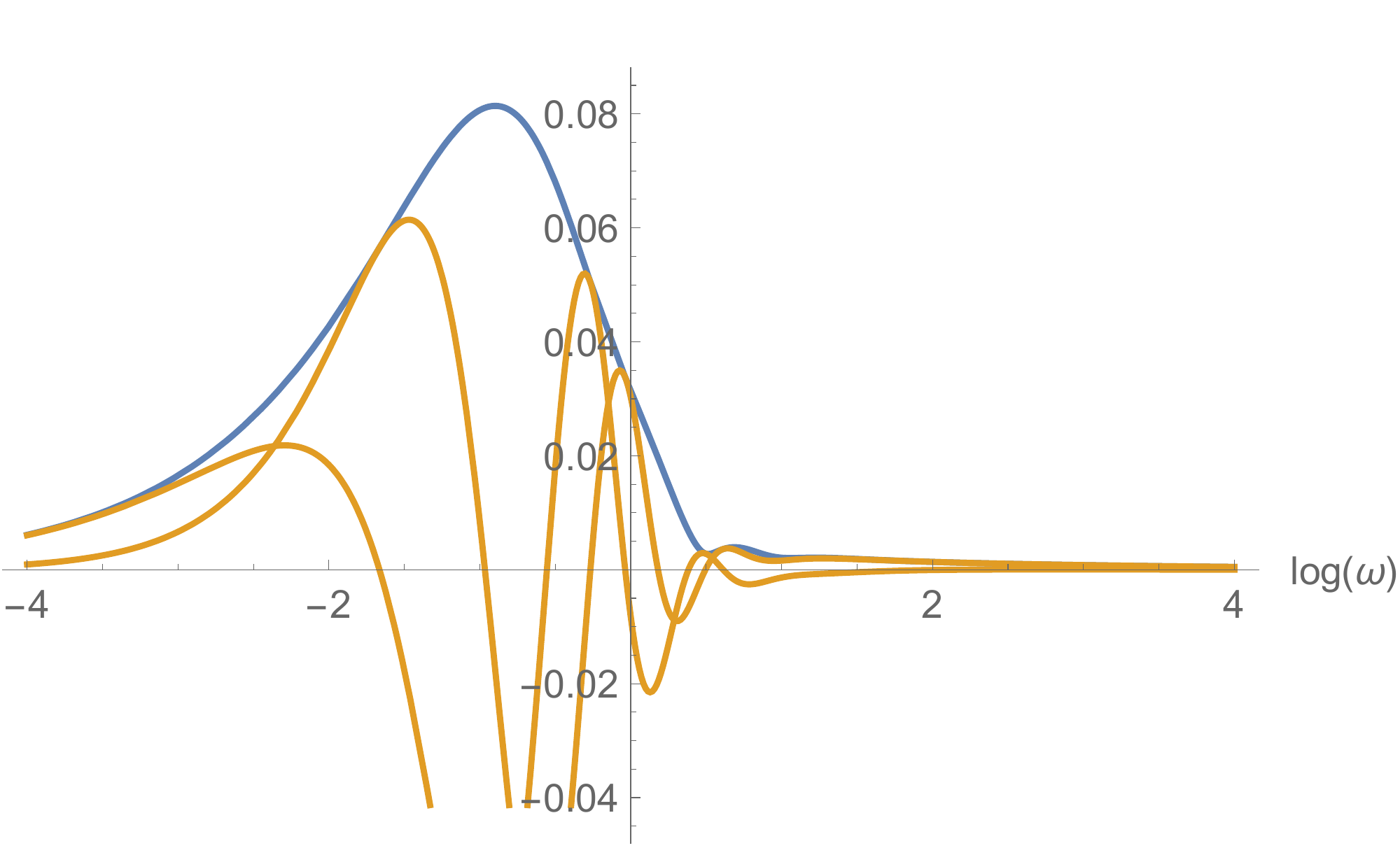}
    \caption{The plot of absolute value (blue), real and imaginary parts (orange) of $G(\omega,1)$ for the five dimensional parabolic case.}
    \label{5dpara2}
\end{figure}
From Figure \ref{5dpara2} in five dimensions we might note to the tiny value of $\abs{G}$ for a specific value of $\omega$. This may be related to the only possible solution ray, but numerical accuracy is insufficient to assess it with certainty. It is given by 
\begin{eqaed}
   \abs{G} \sim 0.006,\quad   \omega \sim 1.65\ .
\end{eqaed}
Note that remarks about the higher dimensional parabolic solutions are given in \cite{Hatefi:2020jdr}. On the other hand three different solutions for the hyperbolic case in four dimensions were determined in \cite{Antonelli:2019dqv}. These three solutions denoted by $\alpha$, $\beta$ and $\gamma$ are summarized in Table~\ref{table:4dhyperbolic}.
\begin{table}
\begin{center}
\tablespacing
\begin{tabular}{|c|c|c|c|}
    \hline
     \multicolumn{4}{|c|}{\textbf{4d hyperbolic}}\\ \hline
      & $\omega$ & $\Im f(0)$  & $z_+$\\ 
      $\alpha$ & $1.362$ & $0.708$ & $1.440$\\
     $\beta$ & $1.003$ & $0.0822$ & $3.29$ \\
     $\gamma$ & $0.541$ & $0.0059$ & $8.44$\\\hline
\end{tabular}
\tablespacing
\end{center}
\caption{The three different solutions in  4d hyperbolic case were found in~\cite{Antonelli:2019dqv}.}
\label{table:4dhyperbolic}
\end{table}

Making use of the root-finding procedure, we also identify a fourth solution $\delta$ for the four-dimensional hyperbolic class (with accuracy less than the other solutions and $G \sim 10^{-5}$) whose parameters are given by
\begin{eqaed}
    \omega & =0.6404, \quad  \Im f(0) = 0.0015, \quad  z_+ &= 19.2923
\end{eqaed} where the graphical representation can be seen in \cite{Hatefi:2020jdr}.
 
 \section{Perturbative analysis}
Here we derive the perturbation equations in general dimensions. We will apply our method for the parabolic case, but it could be taken as an extensive method which holds for all matter content as well.
Note that we have taken some of the steps from~\cite{Hamade:1995jx} while with some algebraic calculations we are able to remove $u(t,r)$ and its derivative from the actual computations \footnote{ Similar perturbations of spherically symmetric background solutions for Horava Gravity were also explored in ~\cite{Ghodsi_2010}.}. We perturb the exact solutions $h_0$ (where $h$ denotes either $b$, $u$ or $f$) found in Section~\ref{sec:unperturbed} according to 

\begin{equation}
 h(t,r) = h_0(z) + \varepsilon \,h_1(t,r)
\end{equation}

where $\varepsilon$ is a small number. If we expand all the equations in powers of $\varepsilon$, then the zeroth order part gives rise to background equations already studied in Section~\ref{sec:unperturbed} and the linearized equations for the perturbations $h_1(t,r)$ are related to linear terms in $\varepsilon $. Let us consider the perturbations of the form

\begin{eqnarray}\label{eq:generic_perturbation_ansatz}
h(z,t) = h_{0}(z) + \varepsilon (-t)^{- \kappa} h_{1}(z) \,,
\label{decay}
\end{eqnarray}

One finds the spectrum of $\kappa$ by solving the equations for $h_1(z)$ and indeed the general solution to the first-order equations is obtained with the linear combination of these modes. We want to find the mode $\kappa^*$ with largest real part (assuming a growing mode for $t \rightarrow 0$, i.e $\Re \kappa > 0$) that is related to the Choptuik exponent by ~\cite{KHA,MA,Hirschmann:1995qx} 
\begin{eqnarray}\label{eq:chopkappa}
\gamma = \frac{1}{\Re \kappa^*}
\end{eqnarray}
Note that just like the four dimensional elliptic case, for simplicity we consider only real modes $\kappa^*$. It can be shown that the values $\kappa = 0$ and $\kappa = 1$ are gauge modes with respect to global $U(1)$ re-definitions of $f$ and time translations respectively, see Section 3.1.1 of \cite{Antonelli:2019dqv}. These modes should be excluded from the computations.
\subsection{Linearized equations of motion in any dimension for the parabolic class}

 Let us apply this program to the parabolic case and explore all the linearized perturbations in arbitrary dimension $d = q+2 \geq 4$. One applies the perturbation ansatz~\eqref{eq:generic_perturbation_ansatz} to all the functions $u$, $b$, $\tau$ as

\begin{equation}\label{eq:pertu}
    u(t,r) = u_0(z) + \varepsilon \, (-t)^{-\kappa} u_1(z) \ ,
\end{equation}
\begin{equation}\label{eq:pertb}
    b(t,r) = b_0(z) + \varepsilon\, (-t)^{-\kappa} b_1(z)\ ,
\end{equation}
\begin{equation}\label{eq:pertelliptictau}
    \tau(t,r) = f(z) + \omega \log(-t)\ ,
\end{equation}
\begin{equation}\label{eq:pertf}
    f(t,r) \equiv f_0(z) + \varepsilon (-t)^{-\kappa} f_1(z)\ .
\end{equation}

One calculates the Ricci tensor for the following metric 
\begin{equation}
 ds^2 = (1+u)( -b^2 dt^2 + dr^2) + r^2 d\Omega_q^2    \; ,
\end{equation}
where $b$ and $u$ should be replaced by the perturbed quantities~\eqref{eq:pertu},~\eqref{eq:pertb}.
The zeroth-order and first-order parts of the Ricci tensor are obtained from
\begin{equation}
    R^{(0)}_{ab} = \lim_{\varepsilon\rightarrow 0} R_{ab}(\varepsilon)\,,
\end{equation}
\begin{equation}
    R^{(1)}_{ab} = \lim_{\varepsilon\rightarrow 0} \dv{R_{ab}(\varepsilon)}{\varepsilon}\,.
\end{equation}
Likewise one does the same for the matter content, applying the axion-dilaton perturbations~\eqref{eq:pertelliptictau},~\eqref{eq:pertf} so that
\begin{equation}
    \tilde{T}^{(0)}_{ab} = \lim_{\varepsilon\rightarrow 0} \tilde{T}_{ab}(\varepsilon)\,,
\end{equation}
\begin{equation}
    \tilde{T}^{(1)}_{ab} = \lim_{\varepsilon\rightarrow 0} \dv{\tilde{T}_{ab}(\varepsilon)}{\varepsilon}\,.
\end{equation}
The Einstein field equations should be held order by order hence
\begin{equation}
    R^{(0)}_{ab} = \tilde{T}^{(0)}_{ab}\,,\quad R^{(1)}_{ab} = \tilde{T}^{(1)}_{ab} \ .
\end{equation}
We now use some of the above equations to remove $u(t,r)$ and its derivatives from the other equations. Indeed by using $R^{(0)}_{tr} = \tilde{T}^{(0)}_{tr}$ we remove $u_0'(z)$, $R^{(0)}_{ij} = \tilde{T}^{(0)}_{ij} = 0$ eliminates $u_0(z)$ (where $i,j$ denote indices on the $(d-2)$-sphere), $R^{(1)}_{tr} = \tilde{T}^{(1)}_{tr}$ also  removes $u_1'(z)$, where eventually $R^{(1)}_{ij} = \tilde{T}^{(1)}_{ij} = 0$ is used to actually remove $u_1(z)$.  From now on we also remove the
 $z$ argument of all functions, so that
 
\begin{align}
      \frac{q u_0'}{2(1+u_0)} &=  \frac{\omega(f_0'+\bar f_0')-2z\bar f_0'f_0'
      }{(f_0-\bar f_0)^2} \, ,\\
    u_0 &= \frac{zb_0'}{(q-1)b_0} \,, \\
    u_1& =-\frac{(q-1)b_1u_0-zb_1'}{(q-1)b_0}\,.
\end{align}

Since the final form of $u_1'(z)$ is complicated we will not write it here. Therefore, $u_0,u_0',u_1,u_1'$  are completely  expressed for all equations in terms of other functions.
Using the following combination of temporal and radial equations of motion 
\begin{equation}
    C(\varepsilon) \equiv R_{tt} + b^2\,R_{rr} - \tilde{T}_{tt} - b^2\, \tilde{T}_{rr} = 0\,,
\end{equation}

we also remove the first derivative terms in $b(t,r)$. 
Indeed we recover the zeroth-order equation as follows

\begin{eqaed} \label{eq:b0prime}
    b_0' = -\frac{2 \left(\left(z^2-b_0^2\right) f_0' \left(z \bar{f_0}'-\omega\right)+\omega \left(\left(b_0^2-z^2\right) \bar{f_0}'+\omega z\right)\right)}{q b_0 (f_0-\bar{f_0})^2} \ ,
\end{eqaed}

where the overbar on $f_0$ denotes complex conjugation.
In the same way the first correction is defined by

\begin{equation}
    \eval{\dv{C(\varepsilon)}{\varepsilon}}_{\varepsilon=0} =0 \ ,\end{equation}
    
which is an equation relating $b_1'$ to $b_0$, $b_0'$ $f_0$, $f_0'$, $f_0''$, and to the other perturbations $b_1$, $f_1$, $f_1'$, and which is really linear in all perturbations. In the parabolic case this equation takes the following form

\begin{eqaed}
    (L_1) b_1^{\prime}&=r ((t-qt)b_0+r b_0^{\prime})\bigg(  -\frac{q t^3 b_1 b_0^{\prime}}{r}+\frac{q\kappa t^4 b_0 b_1 b_0^{\prime}}{r(t-qt)b_0+r^2 b_0^{\prime}} +\frac{4 t^2 b_0 b_1 f_0' \bar f_0'}{s^2}\\
 &
   +\frac{4 t^2 b_0^2  f_0' \bar f_0' (-f_1+\bar f_1)}{s^3}
   +\frac{4 t^2 b_0^2 (f_1-\bar f_1) (t\omega+rf_0')\bar f_0'}{r s^3}\\
 &
   +\frac{2 t^2 b_0^2  f_1' \bar f_0'}{s^2}+\frac{2 t^2 b_0^2  \bar f_0' (\kappa tf_1-rf_1')}{r s^2}+\frac{4 t^2 b_0^2 (f_1-\bar f_1) (t\omega+r\bar f_0')  f_0'}{r s^3}\\
 &
   -\frac{4 (f_1-\bar f_1) (t\omega+rf_0')(t\omega+r\bar f_0') }{ s^3}
  -\frac{2t^2 b_0^2 f_1'(t\omega+r\bar f_0') }{r s^2} \\
 &
  +\frac{2(- \kappa tf_1+rf_1')(t\omega+r\bar f_0') }{ s^2}
  -\frac{4t^2 b_0b_1(t\omega\bar f_0'+f_0'(t\omega+2r\bar f_0')) }{ r s^2} \\
 &
  +\frac{2 t^2 b_0^2  f_0' \bar f_1'}{s^2}
   -\frac{2t^2 b_0^2 (t\omega+r f_0')\bar f_1' }{ r s^2}
   +\frac{2t^2 b_0^2 f_0'(\kappa t\bar f_1-r\bar f_1') }{ r s^2}\\
 &
   +\frac{2(-\kappa t\bar f_1+r\bar f_1') (t\omega+r f_0')}{ s^2}
  \bigg) \; ,
    \label{b1'}\end{eqaed}
with
\begin{equation}
    L_1 = q t^3 b_0 ((\kappa - q+1) t b_0 + r b_0^{\prime}), \quad s= (f_0-\bar f_0) \; .
\end{equation}

 The perturbations are also scale invariant, thus making the coordinate change  $(t,r) \rightarrow (t,z)$, the factors of $t$ cancels out.
We now introduce the perturbation ans\"atze in the $\tau$ equation of motion~\eqref{eq:taueom}. Replacing $b_0'$ according to~\eqref{eq:b0prime}  and solving for $f_0''$,  one recovers the second order background equation for $f_0$,

\begin{eqaed} \label{eq:f0''}
    {q z\left(z^2-b_0^2\right) (f_0-\bar{f_0})^2} f_0''  = &\,\, b_0^2 f_0' \big(2 z f_0' \left(z \bar{f_0}'-\omega\right)-2 q f_0 \left(z f_0'+q \bar{f_0}\right)\\
    &\quad + 2 q z \bar{f_0} f_0'+q^2 f_0^2-2 \omega z \bar{f_0}'+q^2 \bar{f_0}^2\big)\\
    & +z  \Big(2 \omega z \bar{f_0}' \left(\omega-z f_0'\right)+2 q f_0 \big(\left(\omega-z f_0'\right)^2-\bar{f_0} \left(\omega-2 z f_0'\right)\big)\\
    &\quad -2 q \bar{f_0} \left(\omega-z f_0'\right)^2  +q \bar{f_0}^2 \left(\omega-2 z f_0'\right)+q f_0^2 \left(\omega-2 z f_0'\right)\Big)\\
    & +\frac{2 z^3}{b_0^2} \left(\omega-z f_0'\right)^2 \left(\omega-z \bar{f_0}'\right)\ .
\end{eqaed}

Going to first order, the linearized equation for $f_1''$ is

\begin{eqaed} \label{eq:f1doubleprime}
(L_2)f_1''&= \bigg(t^2\kappa(1+\kappa)b_0 f_1-2 t^2b_0b_1b_0' f_0' -t^2b_0^2b_1'f_0'\\
 &+(\kappa tb_1-rb_1')(t\omega +rf_0') -2b_1(t^2 \omega^2+2rt\omega f_0'+(r^2-t^2b_0^2)f_0'^2)m_0
 \\
 & -2b_0(f_1-\bar f_1)(-t^2 \omega^2-2rt\omega f_0'+(-r^2+t^2b_0^2)f_0'^2)m_0^2
 \\
 & 
 -2rt(1+\kappa)b_0f_1'+\frac{qt^3b_0^3f_1'}{r}-t^2b_0^2b_0'f_1'-rb_0'(-t\kappa f_1+rf_1')
 \\
 & +m_0 4b_0\big( t^2b_0b_1f_0'^2+\kappa t f_1(t\omega+rf_0')
+t^2b_0^2 f_0'f_1'-r(t\omega+rf_0')f_1'\big)
 \\
 &
 +\frac{3t^2b_0^2b_1(qtf_0'-rf_0'')}{r}
  -b_1(t^2\omega +2rtf_0'-r^2f_0'')\bigg)\,,
\end{eqaed}

where $L_2= -r^2 b_0+t^2b_0^3$ and $m_0=\frac{1}{f_0-\bar f_0}$. This equation is also scale-invariant. 
By integrating numerically the unperturbed equations, and also substituting $b_1'$ from eq~\eqref{b1'} , we derive the ordinary linear differential equations as follows
\begin{align}
    b_1' & = B_1(b_1,f_1,f_1')\,,\label{eq:equationb1p}\\
    f_1'' & = F_1(b_1,f_1,f_1')\,.\label{eq:equationf1pp}
\end{align}
$B_1$ and $F_1$ are indeed functions linear in the perturbations that have however non-linear dependence on the unperturbed solution. The perturbed equations are also singular at $z=0$ and $b^2(z)=z^2$.
The perturbation equations for hyperbolic case were derived in
\cite{Antonelli:2019dqv} where the modes are explored by finding the  $\kappa$ values that are related to smooth solutions of the perturbed equations~\eqref{eq:equationb1p},~\eqref{eq:equationf1pp}  which need to satisfy the appropriate boundary conditions, which we will now discuss.
\subsection{Boundary conditions for perturbations}
We now turn to boundary conditions needed to solve
Eqs.~\eqref{eq:equationb1p},~\eqref{eq:equationf1pp}. First of all at $z=0$ we rescale the time coordinate, so that $ b_1(0) = 0$, and also using the regularity condition for the axion-dilaton  at $z=0$ we find that $ f_1'(0) = 0$ so that the freedom in $f$ is reduced to $f_1(0)$ which is an unknown complex parameter.  We also demand that at  $z_+$ (we recall that $z_+$ is defined by the equation $b(z_+)=z_+$) all equations and perturbations be regular
so that all the second derivatives $\partial_r^2 f(t,r)$, $\partial_r \partial_t f(t,r)$, $\partial_t^2 f(t,r)$ should be finite as $z\rightarrow z_+$. Hence, $f_0''(z)$ and $f_1''(z)$ are also finite as $z \rightarrow z_+$. For brevity, we introduce $\beta = b_0(z)-z$ and rewrite eqs \eqref{eq:f1doubleprime}-\eqref{eq:f0''} as follows
\begin{align}\label{eq:taylorexpansion_unperturbed}
    f_0''(\beta) & = \frac{1}{\beta} G(h_0) + \mathcal{O}(1)\,,\\ \label{eq:taylorexpansion_linearised}
    f_1''(\beta) & = \frac{1}{\beta^2} \bar{G}(h_0) + \frac{1}{\beta} H(h_0, h_1|\kappa) + \mathcal{O}(1)\,,
\end{align}
where it is understood that  $h_0 = (b_0(z_+),f_0(z_+),f_0'(z_+))$, $h_1 = (b_1(z_+),f_1(z_+),f_1'(z_+))$. The vanishing unperturbed complex constraint is given by $G(h_0)=0$ at $z_+$, and  we checked that it implies 
   $ \bar G(h_0) = 0$  at $z_+$. Hence we are left just with the complex-valued constraint $H(h_0, h_1 | \kappa) = 0$. Finally we solve this constraint for $f_1'(z_+)$ in terms of $f_1(z_+)$, $b_1(z_+)$, $\kappa$ and $h_0$.  Thus this condition does reduce the free parameters at $z_+$ to just a real number $b_1(z_+)$ and a complex $f_1(z_+)$. Finally we will have 6 unknowns including $\kappa$ and the following five-component vector:
\begin{equation}\label{eq:defX}
    X = (\Re f_1(0),\, \Im f_1(0),\, \Re f_1(z_+),\, \Im f_1(z_+),\, b_1(z_+) )
\end{equation}
 We also have the linear ODE's eqs.~\eqref{eq:equationb1p},~\eqref{eq:equationf1pp} whose total real order is five. Let us now briefly explain the numerical procedure. Given a set of boundary conditions $X$, we integrate from $z=0$ to an intermediate point $z_\text{mid}$ and similarly we integrate backwards from $z_+$ to $z_\text{mid}$. Finally  we collect the values of all functions $(b_1, \Re f_1, \Im f_1, \Re f_1', \Im f_1')$ at $z_\text{mid}$ and encode the difference between the two integrations in a ``difference function'' $D(\kappa;X)$. By definition, $D(\kappa; X)$ is linear in $X$ thus it has a representation as a matrix form
\begin{equation}
    D(\kappa; X) = A(\kappa) X
\end{equation}
where $A(\kappa)$ is a $5\times 5$ real matrix depending on $\kappa$. So we need to just find the zeroes of $D(\kappa; X)$ and this can be achieved by evaluating $\det A(\kappa) = 0$. We carry out the root search for the determinant as a function of $\kappa$ where the root with the biggest value will be related to  the Choptuik exponent through eq.~\eqref{eq:chopkappa}. It is worth highlighting one last point: the perturbed equations of motion are singular whenever
the factor
    $W = \left(\kappa+1-q -z \frac{b_0'}{b_0}\right)$ in the denominator vanishes, so that the numerical procedure fails at particular point. We can get an estimate for the values of $\kappa$ giving rise to this singular behaviour as follows, 
\begin{equation}
    0 = \eval{W}_{z = z_+} = \kappa_\text{sing}+1-q- b_0'(z_+)\quad \Rightarrow \quad \kbad = q-1 +b_0'(z_+)\,.
\end{equation}
However, this apparent problem does not affect our evaluation of the critical exponent because in most cases the most relevant mode $\kappa^*$ lies outside that particular failure region.

\section{Results}
\begin{figure}
    \centering
    \includegraphics[width=4in]{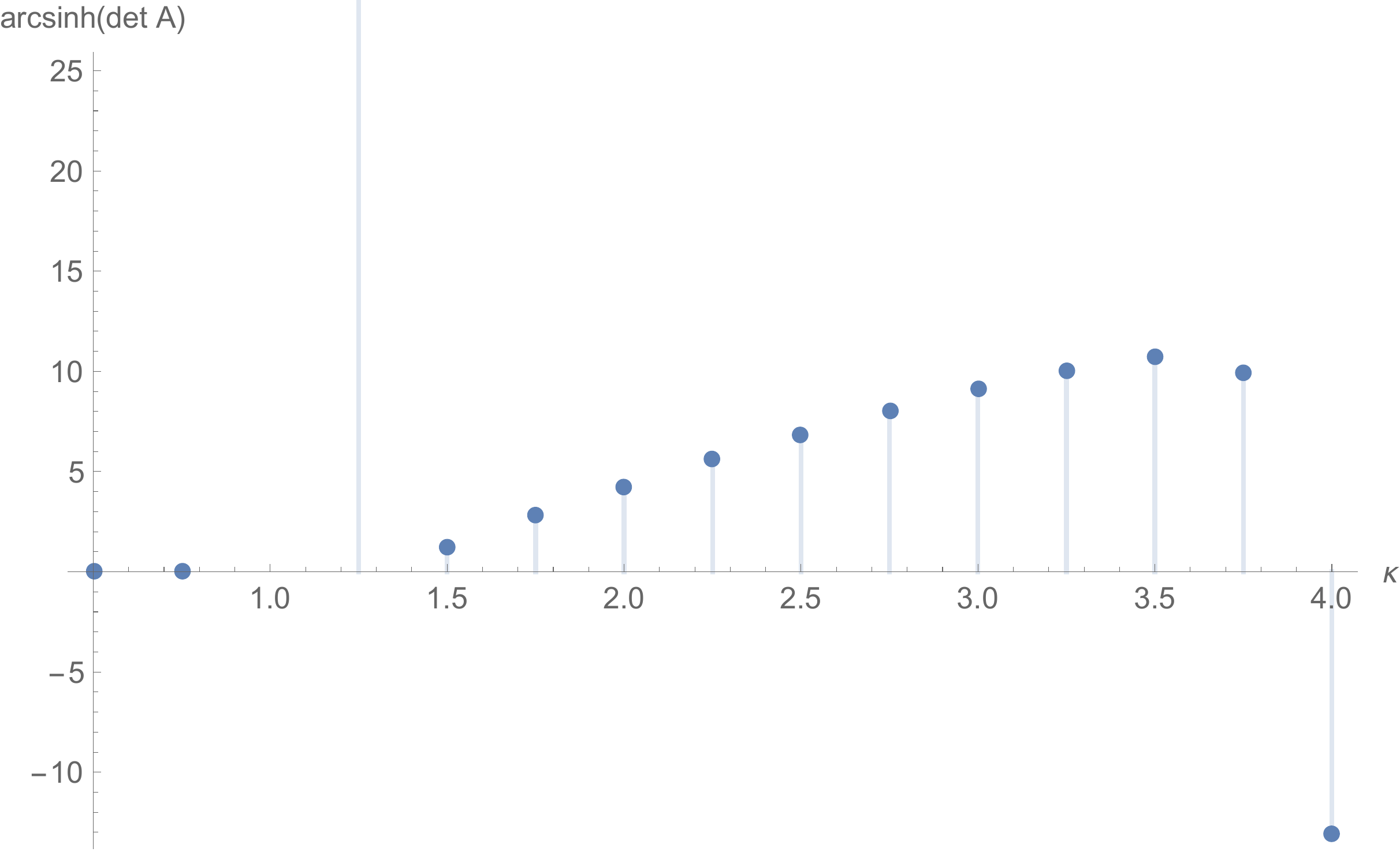}
    \caption{Behaviour of $\operatorname{arcsinh}(\det A(\kappa))$  as a function of $\kappa$ for the unique four-dimensional elliptic critical solution. For clarity, we plot $\operatorname{arcsinh}(\det A(\kappa))$ in order to limit the range of values.}
    \label{fig:choptuik_large}
\end{figure}

 In~\cite{Antonelli:2019dqv} we had already tested the above techniques and were able to derive the critical exponent for the unique four-dimensional elliptic solution. For completeness  we have drawn the behaviours of $\det A(\kappa)$ near the last crossing of the horizontal axis in Figure~\ref{fig:choptuik_large}.  
 The position of the crossing is found to be $ \kappa^*_{4E} \approx 3.7858$, 
that gives rise to the Choptuik exponent $\gamma_{4E} \approx 0.2641$ which is in agreement with \cite{Eardley:1995ns}. Notice that for this solution, $\kbad = 1.224$, and the integration fails around $1 \lesssim \kappa \lesssim 1.4$, which is well below the location of the most relevant mode as it is seen in a range of $\kappa$ values in Figure~\ref{fig:choptuik_large}, where Mathematica was not able to complete the computation of $\det A$.

\begin{figure}
    \centering
    \includegraphics[width=4in]{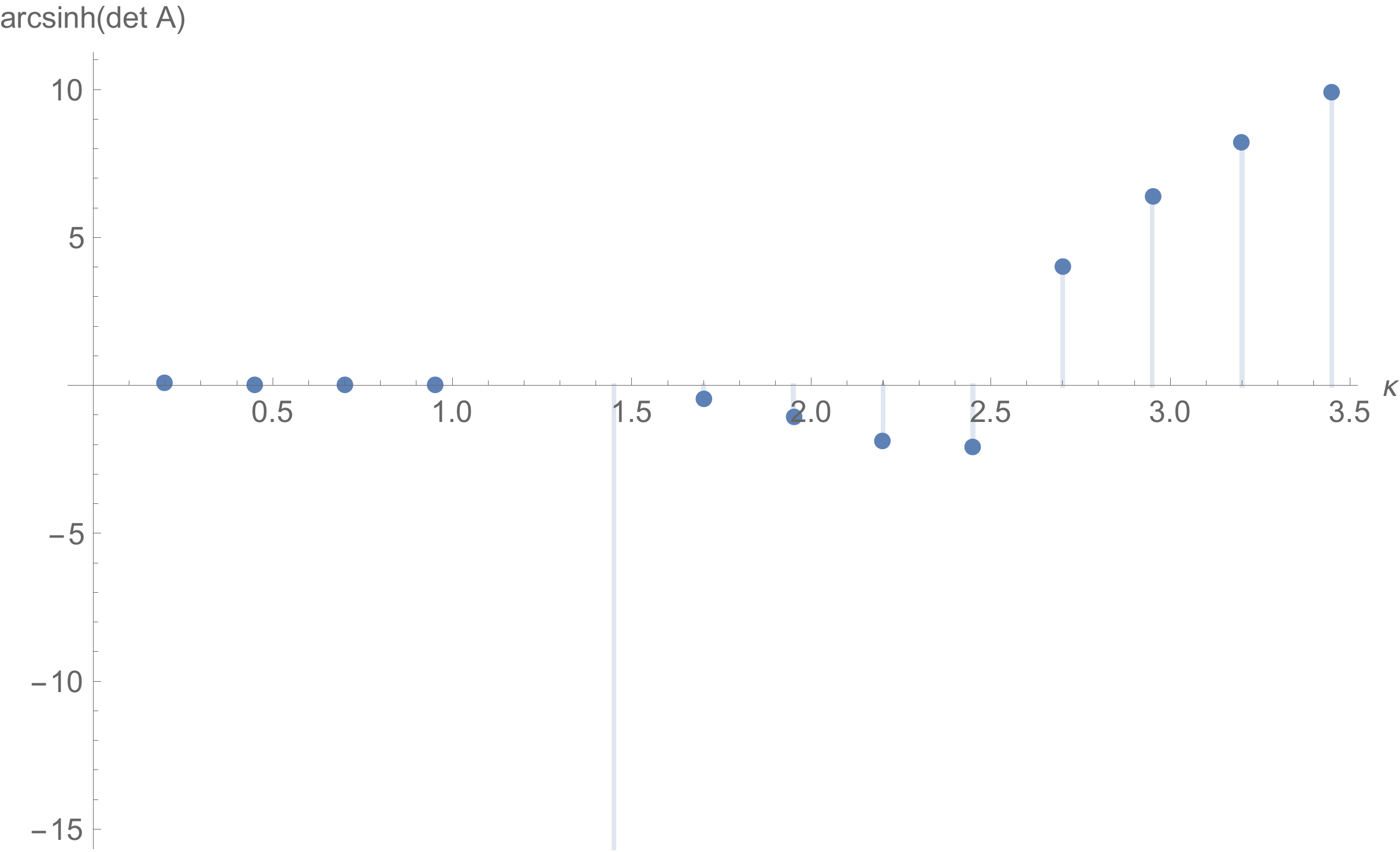}
    \caption{Behaviour of $\operatorname{arcsinh}(\det A(\kappa))$ as a function of $\kappa$ for the fourth soltion of 4 dimensional hyperbolic case, denoted $4H\delta$. For clarity, we plot $\operatorname{arcsinh}(\det A(\kappa))$ in order to limit the range of values.}
    \label{4dHDELCHOP}
\end{figure}

\begin{figure}
    \centering
    \includegraphics[width=4in]{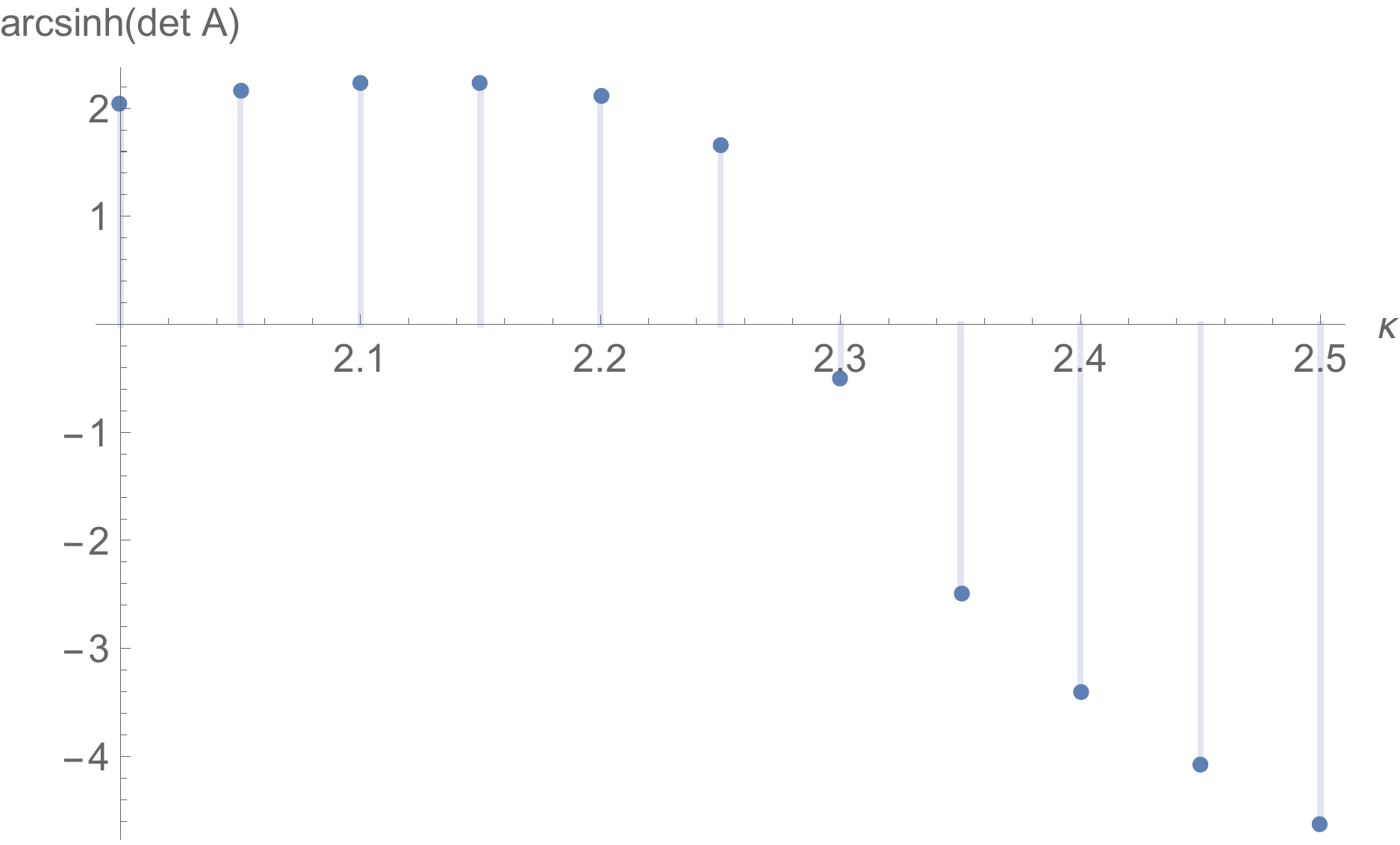}
    \caption{ The behaviour of $\operatorname{arcsinh}(\det A(\kappa))$ for the 4H$\gamma$ solution near the last crossing.}
    \label{fig:4hgamma}
\end{figure}

In the 4d hyperbolic case, there are four branches of solutions that we denote by $4H\alpha$, $4H\beta$, $4H\gamma$ and $4H\delta$ respectively. The Choptuik exponent was not known in the $4H\delta$ case, which is one of the new results of this article. In Figure \ref{4dHDELCHOP} we plot the behaviour of $\det A(\kappa)$ near the last crossing which defines the most relevant mode,
\begin{equation}
    \kappa^*_{4H{\delta}} \approx 2.5456\,,
\end{equation}

so that the Choptuik exponent is

\begin{equation}
   \gamma_{4H{\delta}} \approx 0.393
\end{equation}

 which is different from the Choptuik exponent for the third critical solution $\gamma_{4H{\gamma}}=0.436$ (already found in~\cite{Antonelli:2019dqv}) that is illustrated in \ref{fig:4hgamma}. We collect these results in table~\ref{tab:results333}.

\begin{figure}
    \centering
    \includegraphics[width=4in]{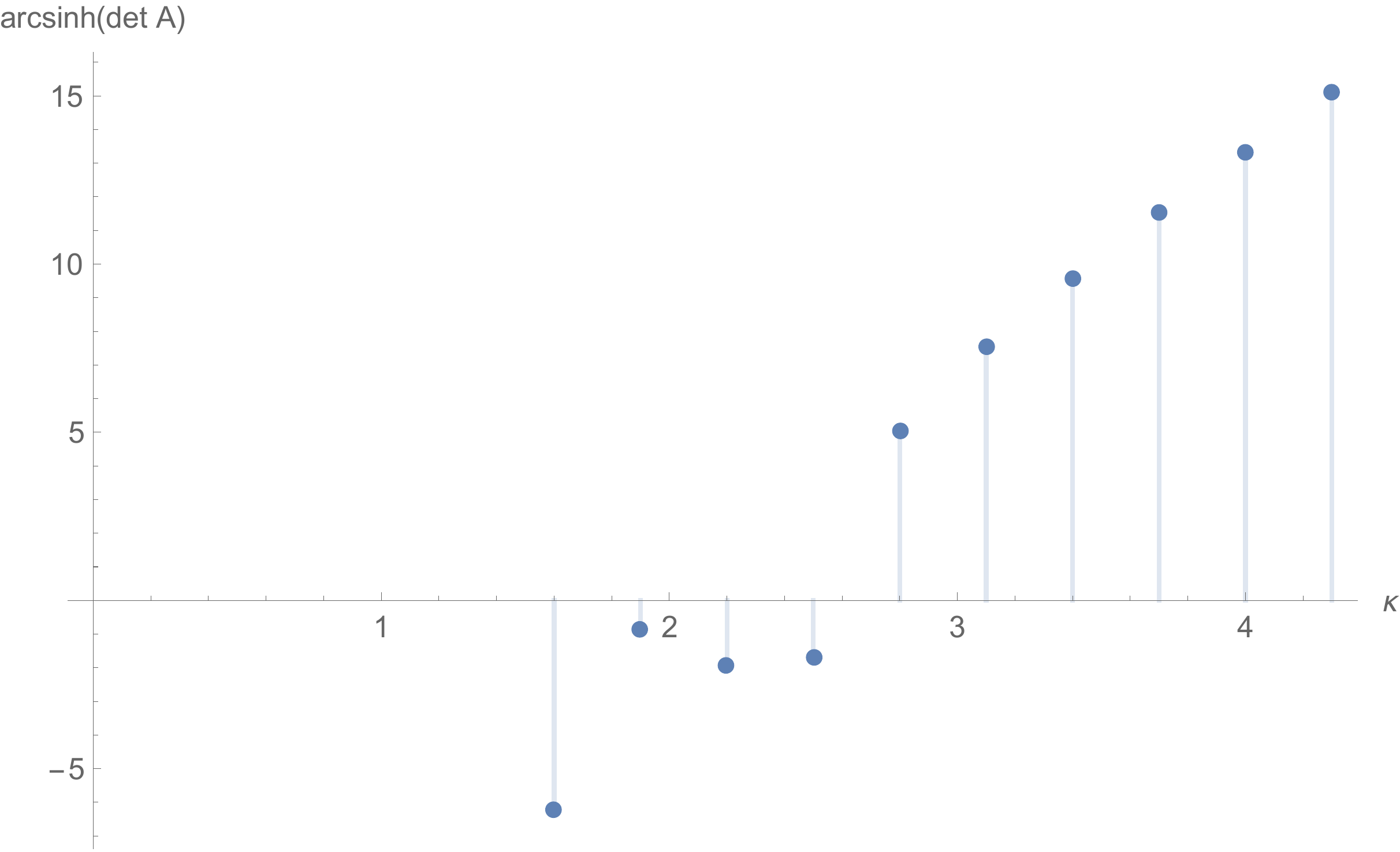}
    \caption{ A zoom on the last crossing of the plot of Figure~\ref{4dHDELCHOP}}
    \label{4HDElta}
\end{figure}

For completeness, we include some other Choptuik exponents in table~\ref{tab:results}. We refer the reader to \cite{Antonelli:2019dqv} for a complete discussion of these other cases.
\begin{table}[H]
\begingroup
\renewcommand*{\arraystretch}{2}
\begin{center}
\begin{tabular}{|c|c|c|c|c|c|c}
\hline
solution & $\kappa^*$ & $\gamma$ &\begingroup \renewcommand*{\arraystretch}{1}\begin{tabular}{@{}c@{}}failure\\region\end{tabular} \endgroup& $\kbad$ & comments \\ \hline \hline
4H$\delta$ & $2.5456$  & $0.393$ & $1-1.58$ & $1.3304$ & see Figure~\ref{4HDElta} \\ \hline
4H$\gamma$ & $2.293$  & $0.436$ & $1-1.5$ & $1.32$ & see Figure~\ref{fig:4hgamma}\\ \hline
\end{tabular}
\end{center}
\endgroup
\caption{Choptuik exponents for the two last branches of solutions of the 4 dimensional hyperbolic class.}
\label{tab:results333}
\end{table}
\begin{table}[ht]
\begingroup
\renewcommand*{\arraystretch}{2}
\begin{center}
\begin{tabular}{|c|c|c|c|c|c|}
\hline
solution & $\kappa^*$ & $\gamma$ &\begingroup \renewcommand*{\arraystretch}{1}\begin{tabular}{@{}c@{}}failure\\region\end{tabular} \endgroup& $\kbad$ & comments \\ \hline \hline
4E & $3.7858$ & $0.2641$ & $1-1.4$ & $1.224$ &  see Figures~\ref{fig:choptuik_large}
\\ \hline
4H$\alpha$ & $1-1.5$ & $0.66-1$ & $1-1.5$ & $1.50$ & $\kappa^*$ likely inside failure region  \\
4H$\beta$ & $1 - 1.55$ & $0.64 - 1$ & $1-1.55$ & $1.4$ &  $\kappa^*$ inside failure region  \\
5E$\alpha$ & $1.186$ & $0.843$ & $2-2.25$ & $2.21$ & \\
5H$\alpha$ & $1.546$ & $0.647$ & $2.1-2.3$ & $2.43$ &    \\ 
\hline
\end{tabular}
\end{center}
\endgroup
\caption{ Partial results for the Choptuik exponents for the 4 and 5 dimensional elliptic and hyperbolic classes.}
\label{tab:results}
\end{table}

\section{Conclusion}

In this article, we have obtained the linear perturbation equations in all classes of solutions of the self-similar collapse solution to the Einstein-axion-dilaton system, including the parabolic case which was not studied previously. The method which we employ is quite generic and could be applied to any matter content in arbitrary dimensions as well. This is certainly a path that we intend to follow in the future.

Through a numerical procedure, we have obtained the fastest growing mode of the perturbations that determine the Choptuik exponent. We have applied this methodology to a particular branch of solutions whose Choptuik exponent was still unknown. Interestingly, we revealed that not only the Choptuik exponent does depend on the spacetime dimension but also it depends on  matter content (which is composed of an axion-dilaton system in this case) as well as the different branches of solutions of self-similar critical collpase. 
Hence, one may conclude that the original conjecture about the universality of Choptuik exponent is not satisfied. However, there might actually exist some universal behaviours hidden in combinations of critical exponents and various other parameters of the theory which have not been taken into account by our current efforts.
\section*{Acknowledgments}
EH would like to thank  R. Antonelli, E. Hirschmann, L. \'Alvarez-Gaum\'e and A. Sagnotti  for useful conversations. This work is supported by INFN (ISCSN4-GSS-PI), by Scuola Normale Superiore, 
and by MIUR-PRIN contract 2017CC72MK003.

\end{document}